\documentclass[twocolumn,showpacs,preprintnumbers,amsmath,amssymb,superscriptaddress]{revtex4-1}

\usepackage{graphicx}
\usepackage{dcolumn}
\usepackage{bm}
\usepackage{color}
\usepackage{soul}

\begin{document}

\title{Comment on ``Optical Imaging of Light-Induced Thermopower in Semiconductors'' {\normalfont[Phys. Rev. Applied {\bf 5}, 024005 (2016)]}}

\author{Y. Apertet}\email{yann.apertet@gmail.com}
\date{\today}

\begin{abstract}
In a recent article [Phys. Rev. Applied {\bf 5}, 024005 (2016)], Gibelli and coworkers proposed a method to determine the thermopower, i.e. the Seebeck coefficient, using photoluminescence measurements. The photoluminescence spectra are used to obtain the local gradients of both the electrochemical potential difference between electron and holes and the temperature of the electron-hole plasma. However, the definition of the thermopower given in that article seems erroneous due to a confusion between the different physical quantities needed to derive this parameter.
\end{abstract} 


\maketitle

\section{Introduction}
The Seebeck coefficient $\alpha$ of a material, also known as thermopower, is a measure of the magnitude of an induced thermoelectric voltage $V_{TE}$ \emph{in response to} a temperature difference across that material $\Delta T$. For sufficiently small temperature difference, there is a linear relation between thermoelectric voltage and the temperature difference. The Seebeck coefficient is the proportionality coefficient between these two quantities obtained under open-circuit condition. It is defined as \cite{Zhou2005}:
\begin{equation}\label{exp}
\alpha = -\frac{V_{TE}}{\Delta T}.
\end{equation}
\noindent Unfortunately, it is not possible to experimentally determine this coefficient for a single material using classical electrical measurements: In practice, one can only evaluate the Seebeck coefficient for a couple of materials. The Seebeck coefficient of a material can thus only be determine if the Seebeck coefficient of the other material composing the thermocouple is known (see, e.g., Ref.~\cite{Handbook}). 
Recently, Gibelli and coworkers reported an analysis of the thermopower induced by a laser beam on a semiconducting material \cite{Gibelli2016}. They proposed to determine Seebeck coefficient from spectrally resolved photoluminescence images. To do so, they extract from these images both temperature and electrochemical potential difference of the electrical carriers inside the sample. These authors claim that such technique allows to get rid of uncertainties linked to classical electrical measurements as it is totally contactless. In the present Comment, we show that their study contains several misleading points: As a result, the quantity derived in Ref.~\cite{Gibelli2016} does not correspond to the traditional Seebeck coefficient. In this Comment, we first demonstrate that the thermopower is ill-defined in Ref.~\cite{Gibelli2016}. Then, we discussed the consistency of the data used by Gibelli and coworkers.

\section{Discussion}
\subsection{On the definition of the Seebeck coefficient}

To compute the thermopower of the sample considered in Ref.~\cite{Gibelli2016}, i.e., an intrinsic quantum-well structure of InGaAsP alloy, Gibelli and coworkers use photoluminescence data as both a thermometer and a voltmeter: For each point of the sample, they gets both the effective temperature $T_{\rm H}$, representative of the electron-hole plasma temperature, and the the quasi Fermi level splitting $\mu_\gamma$. This latter quantity is particularly appealing as it represents the internal voltage of the material, i.e., the maximum open circuit voltage at the edges of the sample \cite{Delamarre2012, Rodiere2014}. This voltage is thus the electrochemical potential difference between the top of the sample made of InGaAs and the bottom of the sample made of InP: The quasi Fermi level splitting $\mu_\gamma$ might consequently be considered as an image of the cross-plane voltage.
From these two quantities, Gibelli and coworkers defines the Seebeck coefficient as:
\begin{equation}\label{defGibelli}
\alpha = \frac{\overrightarrow{\rm grad}~\mu_{\gamma}}{e.\overrightarrow{\rm grad}~T_{\rm H}}
\end{equation}
\noindent where $e$ is the elementary electric charge. This relation is supposed to be the local equivalent of Eq.~(\ref{exp}). However, there are several concerns about this direct identification. First, while one should use the open-circuit voltage, here Gibelli and coworkers use the \emph{variation of} this voltage as they consider the \emph{gradient} of the quasi Fermi level splitting. It is surprising as the physical natures of a quantity and of the gradient of this same quantity are quite different. Secondly, the temperature measurement gives access to the in-plane temperature gradient. Relating the lateral variation of temperature with the variation of the cross-plane voltage is misleading as the former obviously do not give rise to the latter. Note that cross-plane temperature variation is \emph{not} considered in Ref.~\cite{Gibelli2016}. Finally, the appearance of a thermoelectric voltage  under open-circuit condition is associated with the compensation of thermodynamical forces: In this case, the electrochemical potential gradient exactly counterbalances the effects of the imposed temperature gradient (or vice-versa). This behavior is a typical illustration of Le Chatelier's principle. In the experiment conducted by Gibelli and coworkers, both gradients do not compensate each other since they both stem from the laser beam. It thus seems inappropriate to consider that the quasi Fermi level splitting is a thermoelectric voltage.

A possible explanation to the use of Eq.~(\ref{defGibelli}) by Gibelli and coworkers might be its similarity with the classical definition of the Seebeck coefficient given for example by Wood:
\begin{equation}\label{def}
\alpha = \frac{\overrightarrow{\rm grad}~\mu}{e.\overrightarrow{\rm grad}~T}
\end{equation}
\noindent where $\mu$ is the electrochemical potential of the carriers and $T$ is the local temperature \cite{Wood1988}. This hypothesis is supported by the fact that the authors of Ref.~\cite{Gibelli2016} mistakenly describe $\mu_\gamma$ as the electrochemical potential, and no longer as the electrochemical potential \emph{difference}, when introducing Eq.~(\ref{defGibelli}) (see also the caption of Fig.~2(c) in Ref.~\cite{Gibelli2016}). However, the comparison between these two definitions, Eq.~(\ref{defGibelli}) and Eq.~(\ref{def}), further questions the validity of the first of them. Indeed, while Eq.~(\ref{def}) applies to materials with a unique type of electrical carriers, the case considered by Gibelli and coworkers involves both electrons and holes. Unfortunately, the thermopower for an ambipolar material, i.e., where charge carriers of both signs are present, is a sum of the relative contributions of each carrier type weighted by their contributions to the electrical conductivity $\sigma$ \cite{Wood1988}. As none of these quantities can be determined with the approach developed in Ref.~\cite{Gibelli2016}, the traditional definition given in Eq.~(\ref{def}) is useless here.

Hence, the quantity identified by Gibelli and coworkers as the thermopower does not correspond to the genuine thermopower of the studied material. To this extent, the method presented in Ref.~\cite{Gibelli2016} cannot be considered as an alternative to the traditional techniques for the Seebeck coefficient determination.

\subsection{On the data consistency}

In addition to confusion associated with the definition of the Seebeck coefficient, it appears that Ref.~\cite{Gibelli2016} contains data inconsistency. It indeed seems that gradients of both electrochemical potential difference $\mu_\gamma$ and temperature $T_H$ demonstrate the same orientation when comparing Fig.~2(c) and Fig.~2(d) of Ref.~\cite{Gibelli2016} as these two quantities possess the same spatial evolution. This fact is however in contradiction with the slope of the curve on Fig.~3(a) of that article: Due to identical gradient orientations, the slope of this curve, and consequently the Seebeck coefficient, should be positive. This inconsistency further questions the validity of the approach proposed in Ref.~\cite{Gibelli2016}.

Moreover, the determination of the electrochemical potential difference $\mu_\gamma$ seems also questionable: The map of $\mu_\gamma$ is displayed for a particular example on Fig.~2(c) of the commented article \cite{Gibelli2016}. Far from the laser beam, the electrochemical potential difference seems to be stable with a value around 0.56~eV. Yet, as this part of the sample should be at equilibrium, one thus expects the electrochemical potential of electrons and the electrochemical potential of holes to be identical, i.e., $\mu_\gamma=0$. It is unclear if the scale associated with Fig.~2(c) has been restricted, omitting the lower range of the scale, without stressing this fact or if the electrochemical potential difference actually does not vanish. In the latter case, it would suggest that the determination of the electrochemical potential difference $\mu_\gamma$ might be erroneous.

\section{Conclusion}
In conclusion, the results obtained by Gibelli and coworkers should be considered with caution as the Seebeck coefficient is ill-defined in their article due to confusions in the definition of this latter quantity. Moreover, some inconsistencies in the data analysis further cast doubt on the validity of the method described in Ref.~\cite{Gibelli2016}.

\end{document}